\documentclass[a4paper,english,aps,showpacs,preprint]{revtex4}
\usepackage[T1]{fontenc}
\usepackage[latin1]{inputenc}
\usepackage{graphicx}

\makeatletter

\providecommand{\tabularnewline}{\\}

\usepackage{babel}
\makeatother
\begin{document}

\preprint{This line only printed with preprint option}

\title{Theoretical Study of Electron-Phonon Interaction in $ZrB_{2}$ and
$TaB_{2}$ }

\author{Prabhakar P. Singh}

\email{ppsingh@phy.iitb.ac.in}

\affiliation{Department of Physics, Indian Institute of Technology, Powai, Mumbai-
400076, India }

\begin{abstract}
Using full-potential, density-functional-based methods we have studied
electron-phonon interaction in $ZrB_{2}$ and $TaB_{2}$ in $P6/mmm$
crystal structure. Our results for phonon density of states and Eliashberg
function show that the electron-phonon coupling in $ZrB_{2}$ is much
weaker than in $TaB_{2}$. In particular, we find that the average
electron-phonon coupling constant $\lambda$ is equal to $0.15$ for
$ZrB_{2}$ and $0.73$ for $TaB_{2}$. The solutions of the isotropic
Eliashberg gap equation indicate no superconductivity for $ZrB_{2}$
but a superconducting transition temperature $T_{c}$ of around $12\, K$
for $TaB_{2}$ with $\mu^{*}\sim0.16$. 
\end{abstract}

\pacs{63.20.Kr,74.25.Kc}

\maketitle

\section{introduction}

The search for superconductivity in the transition-metal diborides
$ZrB_{2}$ and $TaB_{2}$ has proven to be elusive since the first
reports of superconductivity in $ZrB_{2}$ \cite{gasparov} and $TaB_{2}$
\cite{kaczorowski} with superconducting transition temperatures $T_{c}=5.5\, K$
and $T_{c}=9.5\, K$, respectively. The subsequent experimental efforts
\cite{yamamoto,rosner1} to make $ZrB_{2}$ and $TaB_{2}$ superconducting
by applying external pressure and/or hole doping have also proven
to be futile. Such an outcome for $ZrB_{2}$ and $TaB_{2}$ seems
somewhat surprising given that under similar conditions $Nb_{x}B_{2}$
\cite{yamamoto} and $Mo_{0.96}Zr_{0.04}B_{2}$ \cite{muzzy} are
found to superconduct with $T_{c}=8.5\, K$ and $T_{C}=5.9\, K$,
respectively.

To be able to understand the superconducting properties or lack thereof
of the transition-metal diborides $ZrB_{2}$ and $TaB_{2}$ within
the conventional BCS theory of superconductivity one needs to know
the electron-phonon interaction in these systems. In particular, the
strength of the electron-phonon interaction in $ZrB_{2}$ and $TaB_{2}$,
as reflected in the average electron-phonon coupling constant $\lambda$,
can be used to predict the possibility of superconductivity in the
diborides. Experiments, based on point-contact spectroscopy \cite{yanson,naidyuk},
indicate $\lambda$ to be less than $0.1$ for both $ZrB_{2}$ and
$TaB_{2}$, and thus preclude the possibility of superconductivity
in these diborides. Here, we like to point out that the experimental
determination of $\lambda$, as carried out in Refs. \cite{yanson,naidyuk},
is more reliable for $ZrB_{2}$ than for $TaB_{2}$. 

Previous theoretical estimates of $\lambda$ \cite{rosner1,rosner},
based on only zone-center sampling of the electron-phonon matrix elements,
also indicate $ZrB_{2}$ and $TaB_{2}$ to have a weak electron-phonon
coupling with $\lambda<0.2$. However, it turns out that in the transition-metal
diborides the estimates of $\lambda$ based on zone-center sampling
of the electron-phonon matrix elements corresponding to the optical
$E_{2g}$ mode is not reliable \cite{pps_plam,pps_nbb2}. Thus, a
detailed analysis of the electron-phonon interaction using a more
representative Brillouin zone sampling of the electron-phonon matrix
elements is clearly needed to better understand the lack of superconductivity
in $ZrB_{2}$ and $TaB_{2}$. The present work is a step in that direction. 

Using density-functional-based methods we have studied (i) the electronic
structure, (ii) the phonon density of states (DOS), (iii) the electron-phonon
interaction, and (iv) the solutions of the isotropic Eliashberg gap
equation for $ZrB_{2}$ and $TaB_{2}$ in $P6/mmm$ crystal structure.
We have calculated the electronic structure of $ZrB_{2}$ and $TaB_{2}$
with experimental lattice constants \cite{rosner1,rosner} $a$ and
$c$, as given in Table I, using full-potential linear muffin-tin
orbital (LMTO) method. For studying the electron-phonon interaction
we used the full-potential linear-response program of Savrasov \cite{savrasov1,savrasov2}
to calculate the dynamical matrices and the Hopfield parameters, which
were then used to calculate the phonon DOS, $F(\omega)$, the electron-phonon
coupling $\lambda$, including the partial $\lambda_{\mathbf{q}}$,
and the Eliashberg function, $\alpha^{2}F(\omega)$, for $ZrB_{2}$
and $TaB_{2}$. Subsequently, we have numerically solved the isotropic
Eliashberg gap equation \cite{allen1,allen2,private1} for a range
of $\mu^{*}$ to obtain the corresponding superconducting transition
temperature $T_{c}.$ 

Based on our calculations, described below, we find that the electron-phonon
coupling in $ZrB_{2}$ is much weaker than in $TaB_{2}$. In particular,
we find that the average electron-phonon coupling constant $\lambda$
is equal to $0.15$ for $ZrB_{2}$ and $0.73$ for $TaB_{2}$. The
solutions of the isotropic Eliashberg gap equation indicate no superconductivity
for $ZrB_{2}$ but a superconducting transition temperature $T_{c}$
of around $12\, K$ for $TaB_{2}$ with $\mu^{*}\sim0.16$.

\section{Theoretical Approach and computational details}

Within Migdal-Eliashberg theory of superconductivity \cite{allen1,allen2},
first-principles calculations of superconducting properties require
the knowledge of (i) the ground-state electronic structure, (ii) the
vibrational spectrum and (iii) the electron-phonon matrix elements
of the solid. The phonon spectrum and the electron-phonon matrix elements
are then used to calculate the Eliashberg functions from which the
superconducting properties of the materials can be calculated. In
particular, by solving the fully anisotropic or the isotropic gap
equation the superconducting transition temperature can be calculated. 

The density-functional theory provides a reliable framework for implementing
from first-principles the Migdal-Eliashberg approach for calculating
the superconducting properties of metals as outlined above. Such an
approach has been implemented by Savrasov using the LMTO formalism
and the linear response method, the details of which are given in
Refs. \cite{savrasov1,savrasov2} . In the following, we briefly outline
the approach used in the present calculations. We follow the notation
used in Refs. \cite{savrasov1,savrasov2}.

We calculate the ground-state electronic structure of the transition-metal
diborides $ZrB_{2}$ and $TaB_{2}$ using the full-potential linear
muffin-tin orbital method in the generalized-gradient approximation
of the density-functional theory. The linear response approach is
used to evaluate the dynamical matrices and the electron-phonon matrix
elements within the density-functional theory as implemented using
the LMTO formalism. It amounts to self-consistently evaluating the
changes in the electronic structure due to atomic displacements associated
with the phonon mode $\omega_{\mathbf{q}\nu}$. Once the changes in
the wave function, density and the effective potential are known,
the phonon density of states\begin{equation}
F(\omega)=\sum_{\mathbf{q}\nu}\delta(\omega-\omega_{\mathbf{q}\nu}),\label{eq:fw}\end{equation}
and the electron-phonon matrix elements 

\begin{equation}
g_{\mathbf{k}+\mathbf{q}j',\mathbf{k}j}^{\mathbf{q}\nu}=<\mathbf{k}+\mathbf{q}j'|\delta^{\mathbf{q}\nu}V_{eff}|\mathbf{k}j>,\label{eq:me}\end{equation}
 can be evaluated. The electron-phonon matrix elements can be interpreted
as the scattering of electron in state $|\mathbf{k}j>$ to state $|\mathbf{k+q}j'>$
due to perturbing potential $\delta^{\mathbf{q}\nu}V_{eff}$, which
arises due to the phonon mode $\omega_{\mathbf{q}\nu}$. Note that
Eq. \ref{eq:me} has to be corrected for the incomplete basis set
as described in Ref. \cite{savrasov1}. From the electron-phonon matrix
elements we can calculate the phonon linewidth $\gamma_{\mathbf{q}\nu}$

\begin{equation}
\gamma_{\mathbf{q}\nu}=2\pi\omega_{\mathbf{q}\nu}\sum_{\mathbf{k}jj'}|g_{\mathbf{k}+\mathbf{q}j',\mathbf{k}j}^{\mathbf{q}\nu}|^{2}\delta(\epsilon_{\mathbf{k}j}-\epsilon_{F})\delta(\epsilon_{\mathbf{k+q}j'}-\epsilon_{F}).\label{eq:gamma}\end{equation}
Occasionally, it is useful to define a phonon mode dependent electron-phonon
coupling $\lambda_{\mathbf{q}\nu}$ as 

\begin{equation}
\lambda_{\mathbf{q}\nu}=\frac{\gamma_{\mathbf{q}\nu}}{\pi N(\varepsilon_{F})\omega_{\mathbf{q}\nu}^{2}},\label{eq:lamq}\end{equation}
where $N(\varepsilon_{F})$ is the electronic density of states at
the Fermi energy. Now we can combine the electronic density of states,
phonon spectrum and the electron-phonon matrix elements to obtain
the Eliashberg function $\alpha^{2}F(\omega)$ defined as 

\begin{equation}
\alpha^{2}F(\omega)=\frac{1}{2\pi N(\epsilon_{F})}\sum_{\mathbf{q}\nu}\frac{\gamma_{\mathbf{q}\nu}}{\omega_{\mathbf{q}\nu}}\delta(\omega-\omega_{\mathbf{q}\nu})\label{eq:a2f}\end{equation}
Finally, the Eliashberg functions can be used to solve the isotropic
gap equation \cite{allen1,allen2,private1}\begin{equation}
\Delta(i\omega_{n})=\sum_{n'}^{|\omega_{n'}|<\omega_{c}}f_{n'}S(n,n')\Delta(i\omega_{n'})\label{eq:dn}\end{equation}
 to obtain the superconducting properties such as the superconducting
transition temperature $T_{c}.$ The function $S(n,n')$ used in the
gap equation is defined by \begin{equation}
S(n,n')\equiv\lambda(n-n')-\mu^{*}-\delta_{nn'}\sum_{n''}s_{n}s_{n''}\lambda(n-n'')\label{eq:sn}\end{equation}
and $f_{n}=1/|2n+1|$ with $s_{n}$ representing the sign of $\omega_{n}.$
The electron-phonon coupling $\lambda(\nu)$ is given by \begin{equation}
\lambda(\nu)=\int_{0}^{\infty}d\omega\alpha^{2}F(\omega)\frac{2\omega}{\omega_{\nu}^{2}+\omega^{2}}.\label{eq:lamn}\end{equation}

Before describing our results in detail, we provide some of the computational
details which are similar to our study of the electron-phonon interaction
in $MgB_{2}$, $NbB_{2}$ and $TaB_{2}$ \cite{pps_nbb2,pps_plam}. 

The charge self-consistent full-potential linear muffin-tin orbital
calculations were carried out with the generalized gradient approximation
for exchange-correlation of Perdew \emph{et al} \cite{perdew1,perdew2}
and 484 $\mathbf{k}$-points in the irreducible wedge of the Brillouin
zone. For $TaB_{2}$, the basis set used consisted of $s,$ $p,$
$d$ and $f$ orbitals at the $Ta$ site and $s,$ $p$ and $d$ orbitals
at the $B$ site. In the case of $ZrB_{2}$, we included $s$, $p$
and $d$ orbitals at the $Zr$ site. In all cases the potential and
the wave function were expanded up to $l_{max}=6$. The muffin-tin
radii for $Ta$, $B$, and $Zr$ were taken to be $2.5,$ $1.66,$
and $2.3$ atomic units, respectively. 

The calculation of dynamical matrices and the Hopfield parameters
for $ZrB_{2}$ and $TaB_{2}$ were carried out using a $6\times6\times6$
grid resulting in $28$ irreducible $\mathbf{q}$-points. The Brillouin
zone integrations for charge self-consistency during linear response
calculations were carried out using a $12\times12\times12$ grid of
$\mathbf{k}$-points. The Fermi surface was sampled more accurately
with a $36\times36\times36$ grid of $\mathbf{k}$-points using the
double grid technique as outlined in Ref. \cite{savrasov2}. In case
of $TaB_{2},$ for a couple of $\mathbf{q}$-points we had to sample
the Fermi surface using a $48\times48\times48$ grid of $\mathbf{k}$-points
to get converged acoustic-mode frequencies. 

\begin{table}

\caption{The experimental lattice constants for $ZrB_{2}$ and $TaB_{2}$
used in the calculations.}

\begin{center}\begin{tabular}{|c|c|c|}
\hline 
alloy&
$a$ (a.u.)&
$c/a$ \tabularnewline
\hline
\hline 
$ZrB_{2}$&
5.992&
1.1142\tabularnewline
\hline 
$TaB_{2}$&
5.826&
1.0522\tabularnewline
\hline
\end{tabular}\end{center}
\end{table}

\section{results and discussion}

In this section we describe the results of our calculations of the
electronic structure, the linear response and the solutions of the
isotropic Eliashberg gap equation for $ZrB_{2}$ and $TaB_{2}.$ Our
results for $ZrB_{2}$ and $TaB_{2}$ are described in terms of (i)
the density of states (DOS), (ii) the phonon density of states $F(\omega)$,
(iii) the Eliashberg function $\alpha^{2}F(\omega)$, and (iv) the
superconducting transition temperature $T_{c}$ obtained from the
solutions of the isotropic Eliashberg gap equation.

\subsection{The Density of States}

The electronic structure of $ZrB_{2}$ \cite{rosner,shein} and $TaB_{2}$
\cite{rosner1,shein,pps_tab2} has been studied previously. Our full-potential
results for the electronic structure of $ZrB_{2}$ and $TaB_{2}$
are in agreement with earlier calculations. For example, in Fig. 1
we show the total densities of states for $ZrB_{2}$ and $TaB_{2}$
calculated at the experimental lattice constants using the full-potential
LMTO method as described earlier. The agreement with earlier calculations
of the densities of states for $ZrB_{2}$ \cite{rosner} and $TaB_{2}$
\cite{rosner1} is excellent. We find that the total DOS at the Fermi
energy is equal to $3.72\, st/Ry$ and $12.52\, st/Ry$ for $ZrB_{2}$
and $TaB_{2}$, respectively. In $ZrB_{2}$, the relatively low DOS
at the Fermi energy leads to a weak electron-phonon coupling in this
system. However, at the Fermi energy the $d$-electrons are present
in substantial amount in both $ZrB_{2}$ and $TaB_{2}$, indicating
a more active role for $Zr$ and $Ta$ atoms in determining the lattice-
dynamical as well as the possible superconducting properties of these
diborides. 

\begin{figure}
\begin{center}\includegraphics[%
  width=7.4cm,
  height=7.4cm,
  angle=270]{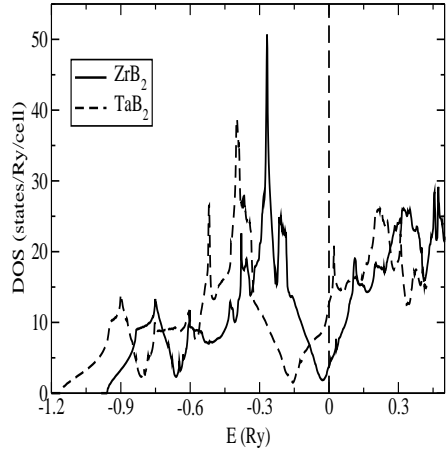}\end{center}

\caption{The total density of states of $ZrB_{2}$ and $TaB_{2}$ calculated
using the full-potential linear muffin-tin orbital method as described
in the text. The vertical dashed line indicates the Fermi energy. }
\end{figure}

\subsection{The Phonon Density of States}

In Fig. 2 we show the phonon DOS $F(\omega)$ of $ZrB_{2}$ and $TaB_{2}$
calculated using the full-potential linear response program as described
earlier. As is clear from Fig. 2, the phonon DOS for both $ZrB_{2}$
and $TaB_{2}$ can be separated into three distinct regions. Based
on our analysis of the eigenvectors, we find that the first region,
with a peak in phonon DOS at $32\, meV$ in $ZrB_{2}$ and $22\, meV$
in $TaB_{2}$, is dominated by the motion of the transition-metal
atoms $Zr$ and $Ta$ respectively. Note that the shift of the first
region in the phonon DOS towards lower frequency for $TaB_{2}$ in
comparison to $ZrB_{2}$ is due to the higher mass of $Ta.$ In the
second region, the phonon DOS around $60-70$ $meV$ in $ZrB_{2}\,(TaB_{2})$
results from the coupled motion of $Zr\,(Ta)$ and the two $B$ atoms.
However, the peaks in the phonon DOS at $78\, meV$ in $ZrB_{2}$
and $75\, meV$ in $TaB_{2}$ correspond to the in-plane $B-B$ motion.
The phonon DOS in the third region, which extends from $94\, meV$
to $102\, meV$ in $ZrB_{2}$ and from $100\, meV$ to $107\, meV$
in $TaB_{2}$, results from the displacements of all the three atoms.
Not surprisingly, the phonon DOS of $ZrB_{2}$ and $TaB_{2}$ are
similar to the phonon DOS of $NbB_{2}$ if we make allowance for the
difference in the masses of the transition-metal atoms \cite{pps_nbb2}. 

\begin{figure}
\begin{center}\includegraphics[%
  width=7.4cm,
  height=7.4cm,
  angle=-90]{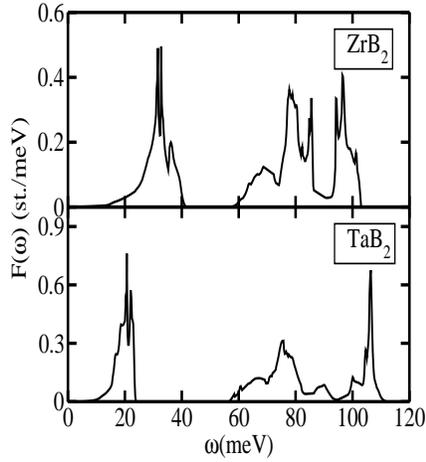}\end{center}

\caption{The phonon density of states $F(\omega)$ of $ZrB_{2}$ and $TaB_{2}$
calculated using the full-potential linear response method as described
in the text.}
\end{figure}

\subsection{The Eliashberg Function}

The main purpose of the present study is to examine the strengths
of the electron-phonon interaction in $ZrB_{2}$ and $TaB_{2}$ in
order to better understand the lack of superconductivity in these
diborides. To this end, we show in Fig. 3 the Eliashberg function
$\alpha^{2}F(\omega)$ of $ZrB_{2}$ and $TaB_{2}$ calculated as
described earlier. A comparison of $\alpha^{2}F(\omega)$ of $ZrB_{2}$
and $TaB_{2}$ shows that the electron-phonon coupling in $ZrB_{2}$
is much weaker than in $TaB_{2}$. In particular, the average electron-phonon
coupling constant $\lambda$ is equal to $0.15$ for $ZrB_{2}$ and
$0.73$ for $TaB_{2}$. 

\begin{figure}
\begin{center}\includegraphics[%
  width=7.4cm,
  height=7.4cm,
  angle=-90]{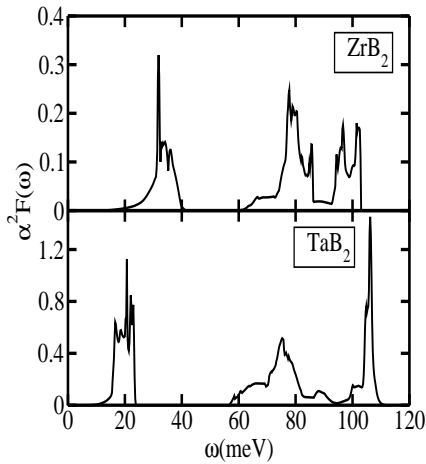}\end{center}

\caption{The Eliashberg function $\alpha^{2}F(\omega)$ of $ZrB_{2}$ and
$TaB_{2}$ calculated using the full-potential linear response method
as described in the text.}
\end{figure}

\begin{table}

\caption{The calculated Hopfield parameter $\eta$, the average electron-phonon
coupling constant $\lambda$, the root mean square $\omega_{rms}$
and the logarithmically averaged $\omega_{ln}$ phonon frequencies
for $ZrB_{2}$ and $TaB_{2}$.}

\begin{center}\begin{tabular}{|c|c|c|c|c|}
\hline 
 alloy&
$\eta\,(mRy/a.u.^{2})$&
$\lambda$&
$\omega_{rms}\,(K)$ &
$\omega_{ln}\,(K)$\tabularnewline
\hline
\hline 
$ZrB_{2}$&
76&
0.15&
734&
585\tabularnewline
\hline 
$TaB_{2}$&
279&
0.73&
593&
356\tabularnewline
\hline
\end{tabular}\end{center}
\end{table}

From Fig. 3, we also see that in the case of $ZrB_{2}$ the phonon
mode with peak at $78\, meV$, which corresponds to the in-plane $B-B$
motion, couples to the electrons more than the phonon mode with peak
at $32\, meV.$ However, in $TaB_{2}$ it is the phonon mode corresponding
to the displacements of the transition-metal atom $Ta$, with a peak
at $22\, meV$, which couples more strongly to the electrons. Such
a change in $\alpha^{2}F(\omega)$ clearly indicates a more active
role for the transition-metal atoms in deciding the normal as well
as the superconducting state (if any) properties of the transition-metal
diborides. Our calculated $\alpha^{2}F(\omega)$ for $ZrB_{2}$ agrees
with the point-contact spectroscopy results of Refs. \cite{yanson,naidyuk}.
However, in our opinion, the experimentally obtained $\alpha^{2}F(\omega)$
for $TaB_{2}$ in Ref. \cite{naidyuk} is underestimated, and it is
in disagreement with the present result. In Table II we have listed
the Hopfield parameter $\eta,$ the electron-phonon coupling constant
$\lambda$, and the various averages of the phonon frequencies for
$ZrB_{2}$ and $TaB_{2}$. 

\begin{figure}
\begin{center}\includegraphics[%
  width=7.4cm,
  height=7.4cm,
  angle=-90]{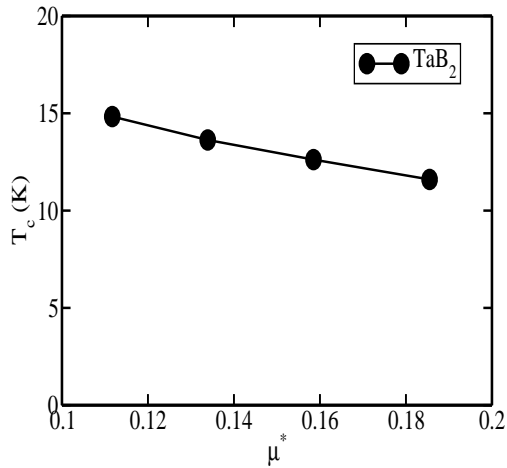}\end{center}

\caption{The superconducting transition temperature $T_{c}$ as a function
of $\mu^{*}$ for $TaB_{2}$ as obtained from the isotropic Eliashberg
gap equation.}
\end{figure}

\subsection{The Superconducting Transition Temperature}

The possibility of superconductivity in $ZrB_{2}$ and $TaB_{2}$
within the present approach can be checked by solving numerically
the isotropic gap equation \cite{allen2,private1} using the calculated
Eliashberg function $\alpha^{2}F(\omega)$. The results of such a
calculation for $TaB_{2}$ are shown in Fig. 4 for a range of values
of $\mu^{*}.$ From Fig. 4 we find that for $\mu^{*}\simeq0.16$ the
$T_{c}$ for $TaB_{2}$ is equal to $\sim12\, K$. A similar calculation
for $ZrB_{2}$ using the calculated Eliashberg function yields a $T_{c}$
of less than $10^{-5}\, K$.

Our results show that $ZrB_{2}$ is unlikely to show superconductivity,
in agreement with experiments \cite{layarovska}. In addition, the
calculated Eliashberg function for $ZrB_{2}$ agrees well with the
corresponding Eliashberg function obtained from point-contact spectroscopy
\cite{yanson,naidyuk}. However, our calculations predict $TaB_{2}$
to superconduct with a $T_{c}\sim12\, K$ but so far no superconductivity
\cite{yamamoto} has been found (except for Ref. \cite{gasparov}).
Although our calculated values of $\alpha^{2}F(\omega)$ and consequently
$\lambda$ can be improved upon by including more \textbf{q}-points
in the linear response calculations but it is unlikely to change the
main conclusions of the present work. Thus, the reasons for the lack
of superconductivity in $TaB_{2}$ need to be explored further. 

\begin{figure}
\begin{center}\includegraphics[%
  width=7.4cm,
  height=7.4cm,
  angle=-90]{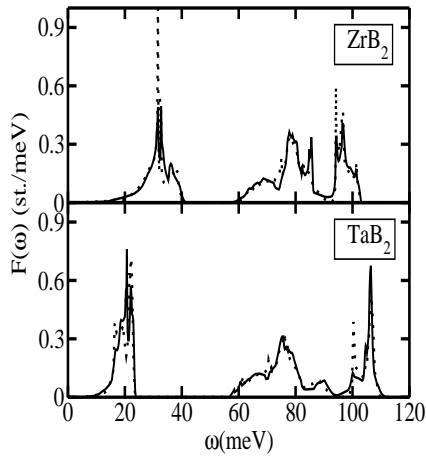}\end{center}

\caption{The phonon density of states $F(\omega)$ of $ZrB_{2}$ and $TaB_{2}$
obtained with $12$ \textbf{q} points (dotted line) and $28$ \textbf{q}
points (solid line) using the full-potential linear response method
as described in the text.}
\end{figure}

\subsection{Convergence of Phonon Density of States, Eliashberg Function and
Superconducting Transition Temperature}

The first-principles study of superconducting properties of solids
as carried out above is computationally demanding. In particular,
the requirement of electronic self-consistency for phonon wave vector
$\mathbf{q}$ and mode $\nu$ makes the computational effort prohibitive
unless one makes a judicious choice of $\mathbf{q}$ vectors. However,
one must ensure that the calculated results are converged with respect
to the number of $\mathbf{q}$ vectors. To demonstrate the convergence
of our calculated phonon density of states $F(\omega),$ Eliashberg
functions $\alpha^{2}F(\omega)$ and the superconducting transition
temperature $T_{c},$ we have carried out two sets of calculations
using the double-grid technique as outlined in Ref. \cite{savrasov1,savrasov2}.
We have used two $\mathbf{q}$ grids (i) a $4\times4\times4$ grid
with $12$ irreducible $\mathbf{q}$-points and (ii) a $6\times6\times6$
grid with $28$ irreducible $\mathbf{q}$-points. All the results
described so far correspond to the second grid with $28$ $\mathbf{q}$-points.
Below we show a comparison of $F(\omega),$ $\alpha^{2}F(\omega)$
and $T_{c}$ calculated using the two grids with $12$ and $28$ $\mathbf{q}$-points,
respectively. 

\begin{figure}
\begin{center}\includegraphics[%
  width=7.4cm,
  height=7.4cm,
  angle=-90]{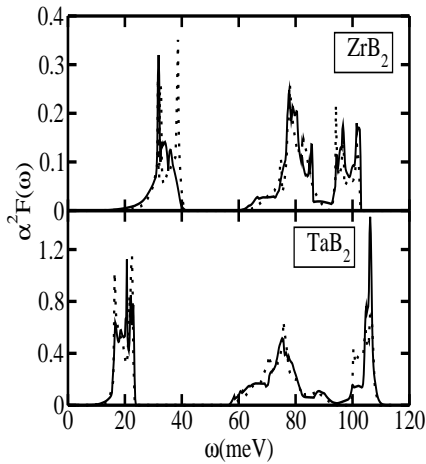}\end{center}

\caption{The Eliashberg function $\alpha^{2}F(\omega)$ of $ZrB_{2}$ and
$TaB_{2}$ obtained with $12$ \textbf{q} points (dotted line) and
$28$ \textbf{q} points (solid line) using the full-potential linear
response method as described in the text.}
\end{figure}

\begin{figure}
\begin{center}\includegraphics[%
  width=7.4cm,
  height=7.4cm,
  angle=-90]{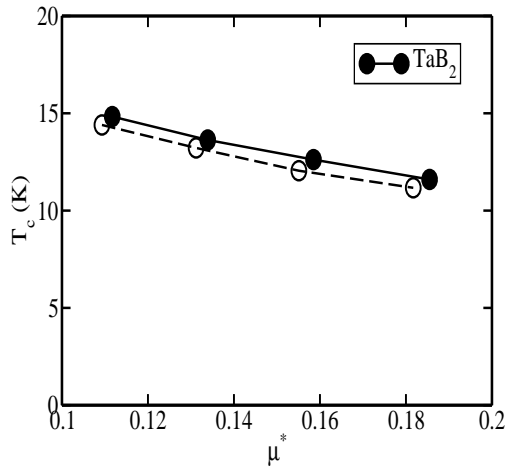}\end{center}

\caption{The superconducting transition temperature $T_{c}$ as a function
of $\mu^{*}$ for $TaB_{2}$ as obtained from the isotropic Eliashberg
gap equation. The dashed and the solid lines correspond to the calculations
done with $12$ and $28$ \textbf{q} points, respectively.}
\end{figure}

In Fig. 5 we compare $F(\omega)$ calculated using the two grids with
$12$ and $28$ $\mathbf{q}$-points, respectively. In general, dynamical
matrices and consequently the phonon spectrum do not require a very
fine $\mathbf{k}$-mesh for convergence. However, the increase in
the number of $\mathbf{q}$-points has understandably reduced the
peak heights at $\omega$ approximately equal to $30\, meV$, $75\, meV$
and $95\, meV$ for $ZrB_{2}$ and $20\, meV$, $70\, meV$ and $100\, meV$
for $TaB_{2}.$ 

The convergence of $\alpha^{2}F(\omega)$ is shown in Fig. 6, where
we have plotted $\alpha^{2}F(\omega)$ calculated using the two grids
with $12$ and $28$ $\mathbf{q}$-points, respectively. For $ZrB_{2}$
the differences are small in magnitude except at around $20\, meV,$
while for $TaB_{2}$ the effects due to small number of \textbf{$\mathbf{q}$}-points
are reduced with the use of the $28$ \textbf{$\mathbf{q}$}-point
grid. We also see that the integrated values such as the electron-phonon
coupling constant $\lambda$ do not change appreciably between the
two sets of calculations ($\lambda$ changes by $\sim0.01$ for $ZrB_{2}$
and $\sim0.005$ for $TaB_{2}$). 

As $F(\omega)$ and $\alpha^{2}F(\omega)$ do not change significantly
between the two sets of calculations, we do not expect a significant
change in the solutions of the isotropic gap equation. Consequently,
the calculated $T_{c}$ should be similar in the two cases. In Fig.
7, we show the $T_{c}$ obtained for the two sets of calculations,
and the difference of less than $1^{0}\, K$ between the two sets
of calculations confirms the convergence of $F(\omega)$, $\alpha^{2}F(\omega)$
and $T_{c}$.

\section{conclusions}

We have studied electron-phonon interaction in $ZrB_{2}$ and $TaB_{2}$
in $P6/mmm$ crystal structure using full-potential, density-functional-based
methods. we find that the electron-phonon coupling in $ZrB_{2}$ is
much weaker than in $TaB_{2}$. In particular, we find that the average
electron-phonon coupling constant $\lambda$ is equal to $0.15$ for
$ZrB_{2}$ and $0.73$ for $TaB_{2}$. The solutions of the isotropic
Eliashberg gap equation indicate no superconductivity for $ZrB_{2}$
but a superconducting transition temperature $T_{c}$ of around $12\, K$
for $TaB_{2}$ with $\mu^{*}\sim0.16$.


\begin{thebibliography}{20}
\expandafter\ifx\csname natexlab\endcsname\relax\def\natexlab#1{#1}\fi
\expandafter\ifx\csname bibnamefont\endcsname\relax
  \def\bibnamefont#1{#1}\fi
\expandafter\ifx\csname bibfnamefont\endcsname\relax
  \def\bibfnamefont#1{#1}\fi
\expandafter\ifx\csname citenamefont\endcsname\relax
  \def\citenamefont#1{#1}\fi
\expandafter\ifx\csname url\endcsname\relax
  \def\url#1{\texttt{#1}}\fi
\expandafter\ifx\csname urlprefix\endcsname\relax\def\urlprefix{URL }\fi
\providecommand{\bibinfo}[2]{#2}
\providecommand{\eprint}[2][]{\url{#2}}

\bibitem[{\citenamefont{Gasparov et~al.}(2001)\citenamefont{Gasparov, Sidorov,
  Zverkova, and Kulakov}}]{gasparov}
\bibinfo{author}{\bibfnamefont{V.~A.} \bibnamefont{Gasparov}},
  \bibinfo{author}{\bibfnamefont{N.~S.} \bibnamefont{Sidorov}},
  \bibinfo{author}{\bibfnamefont{I.~I.} \bibnamefont{Zverkova}},
  \bibnamefont{and} \bibinfo{author}{\bibfnamefont{M.~P.}
  \bibnamefont{Kulakov}}, \bibinfo{journal}{JETP Lett.}
  \textbf{\bibinfo{volume}{73}}, \bibinfo{pages}{532} (\bibinfo{year}{2001}).

\bibitem[{\citenamefont{Kaczorowski et~al.}()\citenamefont{Kaczorowski,
  Zaleski, Zogal, and Klamut}}]{kaczorowski}
\bibinfo{author}{\bibfnamefont{D.}~\bibnamefont{Kaczorowski}},
  \bibinfo{author}{\bibfnamefont{A.~J.} \bibnamefont{Zaleski}},
  \bibinfo{author}{\bibfnamefont{O.~J.} \bibnamefont{Zogal}}, \bibnamefont{and}
  \bibinfo{author}{\bibfnamefont{J.}~\bibnamefont{Klamut}},
  \eprint{cond-mat/0103571}.

\bibitem[{\citenamefont{Yamamoto et~al.}()\citenamefont{Yamamoto, Takao, Masui,
  Izumi, and Tajima}}]{yamamoto}
\bibinfo{author}{\bibfnamefont{A.}~\bibnamefont{Yamamoto}},
  \bibinfo{author}{\bibfnamefont{C.}~\bibnamefont{Takao}},
  \bibinfo{author}{\bibfnamefont{T.}~\bibnamefont{Masui}},
  \bibinfo{author}{\bibfnamefont{M.}~\bibnamefont{Izumi}}, \bibnamefont{and}
  \bibinfo{author}{\bibfnamefont{S.}~\bibnamefont{Tajima}},
  \eprint{cond-mat/0208331}.

\bibitem[{\citenamefont{Rosner et~al.}(2001)\citenamefont{Rosner, Pickett,
  Dreschsler, Handstein, G.Behr, Fuchs, Nevkov, Muller, and
  Eschring}}]{rosner1}
\bibinfo{author}{\bibfnamefont{H.}~\bibnamefont{Rosner}},
  \bibinfo{author}{\bibfnamefont{W.~E.} \bibnamefont{Pickett}},
  \bibinfo{author}{\bibfnamefont{S.~L.} \bibnamefont{Dreschsler}},
  \bibinfo{author}{\bibfnamefont{A.}~\bibnamefont{Handstein}},
  \bibinfo{author}{\bibnamefont{G.Behr}},
  \bibinfo{author}{\bibfnamefont{G.}~\bibnamefont{Fuchs}},
  \bibinfo{author}{\bibfnamefont{K.}~\bibnamefont{Nevkov}},
  \bibinfo{author}{\bibfnamefont{K.}~\bibnamefont{Muller}}, \bibnamefont{and}
  \bibinfo{author}{\bibfnamefont{H.}~\bibnamefont{Eschring}},
  \bibinfo{journal}{Phys.\ Rev. B} \textbf{\bibinfo{volume}{64}},
  \bibinfo{pages}{144516} (\bibinfo{year}{2001}).

\bibitem[{\citenamefont{Muzzy et~al.}()\citenamefont{Muzzy, Avdeev, Lawes,
  Haas, Zandbergen, Ramirez, Jorgensen, and Cava}}]{muzzy}
\bibinfo{author}{\bibfnamefont{L.~E.} \bibnamefont{Muzzy}},
  \bibinfo{author}{\bibfnamefont{M.}~\bibnamefont{Avdeev}},
  \bibinfo{author}{\bibfnamefont{G.}~\bibnamefont{Lawes}},
  \bibinfo{author}{\bibfnamefont{M.~K.} \bibnamefont{Haas}},
  \bibinfo{author}{\bibfnamefont{H.~W.} \bibnamefont{Zandbergen}},
  \bibinfo{author}{\bibfnamefont{A.~P.} \bibnamefont{Ramirez}},
  \bibinfo{author}{\bibfnamefont{J.~D.} \bibnamefont{Jorgensen}},
  \bibnamefont{and} \bibinfo{author}{\bibfnamefont{R.~J.} \bibnamefont{Cava}},
  \eprint{cond-mat/0206006}.

\bibitem[{\citenamefont{Yanson et~al.}()\citenamefont{Yanson, Naidyuk,
  Kvitnitskaya, Fisun, Bobrov, Chubov, Ryabovol, Behr, Kang, Choi
  et~al.}}]{yanson}
\bibinfo{author}{\bibfnamefont{I.~K.} \bibnamefont{Yanson}},
  \bibinfo{author}{\bibfnamefont{Y.~G.} \bibnamefont{Naidyuk}},
  \bibinfo{author}{\bibfnamefont{O.~E.} \bibnamefont{Kvitnitskaya}},
  \bibinfo{author}{\bibfnamefont{V.~V.} \bibnamefont{Fisun}},
  \bibinfo{author}{\bibfnamefont{N.~L.} \bibnamefont{Bobrov}},
  \bibinfo{author}{\bibfnamefont{P.~N.} \bibnamefont{Chubov}},
  \bibinfo{author}{\bibfnamefont{V.~V.} \bibnamefont{Ryabovol}},
  \bibinfo{author}{\bibfnamefont{G.}~\bibnamefont{Behr}},
  \bibinfo{author}{\bibfnamefont{W.~N.} \bibnamefont{Kang}},
  \bibinfo{author}{\bibfnamefont{E.-M.} \bibnamefont{Choi}},
  \bibnamefont{et~al.}, \eprint{cond-mat/0212599}.

\bibitem[{\citenamefont{Naidyuk et~al.}(2002)\citenamefont{Naidyuk,
  Kvitnitskaya, Yanson, Drechsler, Behr, and Otani}}]{naidyuk}
\bibinfo{author}{\bibfnamefont{Y.~G.} \bibnamefont{Naidyuk}},
  \bibinfo{author}{\bibfnamefont{O.~E.} \bibnamefont{Kvitnitskaya}},
  \bibinfo{author}{\bibfnamefont{I.~K.} \bibnamefont{Yanson}},
  \bibinfo{author}{\bibfnamefont{S.-L.} \bibnamefont{Drechsler}},
  \bibinfo{author}{\bibfnamefont{G.}~\bibnamefont{Behr}}, \bibnamefont{and}
  \bibinfo{author}{\bibfnamefont{S.}~\bibnamefont{Otani}},
  \bibinfo{journal}{Phys. Rev. B} \textbf{\bibinfo{volume}{66}},
  \bibinfo{pages}{140301} (\bibinfo{year}{2002}).

\bibitem[{\citenamefont{Rosner et~al.}(2002)\citenamefont{Rosner, An, Pickett,
  and Drechsler}}]{rosner}
\bibinfo{author}{\bibfnamefont{H.}~\bibnamefont{Rosner}},
  \bibinfo{author}{\bibfnamefont{J.~M.} \bibnamefont{An}},
  \bibinfo{author}{\bibfnamefont{W.~E.} \bibnamefont{Pickett}},
  \bibnamefont{and} \bibinfo{author}{\bibfnamefont{S.-L.}
  \bibnamefont{Drechsler}}, \bibinfo{journal}{Phys.\ Rev. B}
  \textbf{\bibinfo{volume}{66}}, \bibinfo{pages}{024521}
  (\bibinfo{year}{2002}).

\bibitem[{\citenamefont{Singh}(2003{\natexlab{a}})}]{pps_plam}
\bibinfo{author}{\bibfnamefont{P.~P.} \bibnamefont{Singh}},
  \bibinfo{journal}{Phys. Rev. B} \textbf{\bibinfo{volume}{67}},
  \bibinfo{pages}{132511} (\bibinfo{year}{2003}{\natexlab{a}}).

\bibitem[{\citenamefont{Singh}(2003{\natexlab{b}})}]{pps_nbb2}
\bibinfo{author}{\bibfnamefont{P.~P.} \bibnamefont{Singh}},
  \bibinfo{journal}{Solid State Commun.} \textbf{\bibinfo{volume}{125/6}},
  \bibinfo{pages}{323} (\bibinfo{year}{2003}{\natexlab{b}}).

\bibitem[{\citenamefont{Savrasov}(1996)}]{savrasov1}
\bibinfo{author}{\bibfnamefont{S.~Y.} \bibnamefont{Savrasov}},
  \bibinfo{journal}{Phys. Rev. B} \textbf{\bibinfo{volume}{54}},
  \bibinfo{pages}{16470} (\bibinfo{year}{1996}).

\bibitem[{\citenamefont{Savrasov and Savrasov}(1996)}]{savrasov2}
\bibinfo{author}{\bibfnamefont{S.~Y.} \bibnamefont{Savrasov}} \bibnamefont{and}
  \bibinfo{author}{\bibfnamefont{D.~Y.} \bibnamefont{Savrasov}},
  \bibinfo{journal}{Phys. Rev. B} \textbf{\bibinfo{volume}{54}},
  \bibinfo{pages}{16487} (\bibinfo{year}{1996}).

\bibitem[{\citenamefont{Allen and Dynes}(1975)}]{allen1}
\bibinfo{author}{\bibfnamefont{P.~B.} \bibnamefont{Allen}} \bibnamefont{and}
  \bibinfo{author}{\bibfnamefont{R.~C.} \bibnamefont{Dynes}},
  \bibinfo{journal}{Phys.\ Rev. B} \textbf{\bibinfo{volume}{12}},
  \bibinfo{pages}{905} (\bibinfo{year}{1975}).

\bibitem[{\citenamefont{Allen and Mitrovic}(1982)}]{allen2}
\bibinfo{author}{\bibfnamefont{P.}~\bibnamefont{Allen}} \bibnamefont{and}
  \bibinfo{author}{\bibfnamefont{B.}~\bibnamefont{Mitrovic}},
  p.~\bibinfo{pages}{1} (\bibinfo{year}{1982}).

\bibitem[{\citenamefont{Allen}()}]{private1}
\bibinfo{author}{\bibfnamefont{P.~B.} \bibnamefont{Allen}}, \eprint{private
  communication}.

\bibitem[{\citenamefont{Perdew and Wang}(1992)}]{perdew1}
\bibinfo{author}{\bibfnamefont{J.~P.} \bibnamefont{Perdew}} \bibnamefont{and}
  \bibinfo{author}{\bibfnamefont{Y.}~\bibnamefont{Wang}},
  \bibinfo{journal}{Phys.\ Rev. B} \textbf{\bibinfo{volume}{45}},
  \bibinfo{pages}{13244} (\bibinfo{year}{1992}).

\bibitem[{\citenamefont{Perdewa et~al.}(1996)\citenamefont{Perdewa, Burke, and
  Ernzerhof}}]{perdew2}
\bibinfo{author}{\bibfnamefont{J.~P.} \bibnamefont{Perdewa}},
  \bibinfo{author}{\bibfnamefont{K.}~\bibnamefont{Burke}}, \bibnamefont{and}
  \bibinfo{author}{\bibfnamefont{M.}~\bibnamefont{Ernzerhof}},
  \bibinfo{journal}{Phys.\ Rev. Lett.} \textbf{\bibinfo{volume}{77}},
  \bibinfo{pages}{3865} (\bibinfo{year}{1996}).

\bibitem[{\citenamefont{Shein and Ivanovskii}()}]{shein}
\bibinfo{author}{\bibfnamefont{I.~R.} \bibnamefont{Shein}} \bibnamefont{and}
  \bibinfo{author}{\bibfnamefont{A.~L.} \bibnamefont{Ivanovskii}},
  \eprint{cond-mat/0109445 v2}.

\bibitem[{\citenamefont{Singh}()}]{pps_tab2}
\bibinfo{author}{\bibfnamefont{P.~P.} \bibnamefont{Singh}},
  \eprint{cond-mat/0104580}.

\bibitem[{\citenamefont{Layarovska and Layarovski}(1979)}]{layarovska}
\bibinfo{author}{\bibfnamefont{L.}~\bibnamefont{Layarovska}} \bibnamefont{and}
  \bibinfo{author}{\bibfnamefont{E.}~\bibnamefont{Layarovski}},
  \bibinfo{journal}{J. Less-Common Metals} \textbf{\bibinfo{volume}{67}},
  \bibinfo{pages}{249} (\bibinfo{year}{1979}).

\end{thebibliography}
\end{document}